\documentclass[conference]{IEEEtran}
\IEEEoverridecommandlockouts
\usepackage{cite}
\usepackage{amsmath,amssymb,amsfonts}
\usepackage{algorithmic}
\usepackage{graphicx}
\usepackage{textcomp}
\usepackage{xcolor}
\usepackage{blindtext}
\usepackage{tabularx}
\def\BibTeX{{\rm B\kern-.05em{\sc i\kern-.025em b}\kern-.08em
    T\kern-.1667em\lower.7ex\hbox{E}\kern-.125emX}}
\begin{document}

\newcommand{\etal}{\textit{et al.}}

\title{
Endorsement-Driven Blockchain SSI Framework for Dynamic IoT Ecosystems
}

\author{
\IEEEauthorblockN{
    Guntur Dharma Putra and
    Bagus Rakadyanto Oktavianto Putra
}
\IEEEauthorblockA{
    Universitas Gadjah Mada, Indonesia \\
    gdputra@ugm.ac.id, bagus.rak2002@mail.ugm.ac.id
}
}

\maketitle

\begin{abstract}
Self-Sovereign Identity (SSI) offers significant potential for managing identities in the Internet of Things (IoT), enabling decentralized authentication and credential management without reliance on centralized entities. However, existing SSI frameworks often limit credential issuance and revocation to trusted entities, such as IoT manufacturers, which restricts flexibility in dynamic IoT ecosystems. In this paper, we propose a blockchain-based SSI framework that allows any individual with a verifiable trust linkage to act as a credential issuer, ensuring decentralized and scalable identity management. Our framework incorporates a layered architecture, where trust is dynamically established through endorsement-based calculations and maintained via a hierarchical chain-of-trust mechanism. Blockchain serves as the Verifiable Data Registry, ensuring transparency and immutability of identity operations, while smart contracts automate critical processes such as credential issuance, verification, and revocation. A proof-of-concept implementation demonstrates that the proposed framework is feasible and incurs minimal overheads compared to the baseline, making it well-suited for dynamic and resource-constrained IoT environments.
\end{abstract}


\section{Introduction} 
Self-Sovereign Identity (SSI) provides transformative opportunities for managing identities in the Internet of Things (IoT), offering an alternative to conventional Public Key Infrastructure (PKI) mechanisms for authentication~\cite{heoDecentralised2024}. By decentralizing Identity Management (IdM), SSI facilitates seamless person-to-device and device-to-device authentication, addressing the scalability and security challenges inherent in IoT ecosystems~\cite{badirovaSurvey2023}. In a typical SSI setup for IoT, credential issuance and revocation are predominantly managed by device manufacturers, who act as the primary trusted issuers~\cite{boiDecentralized2024}. However, this reliance on manufacturers imposes significant limitations, particularly in dynamic IoT environments where devices frequently change ownership or operational context~\cite{satybaldyTaxonomy2024}. For instance, a device sold to a new user requires reissuance of credentials, a process constrained by the centralization of credential management~\cite{fedrecheskiSelfSovereign2020}. Furthermore, scenarios involving stolen or compromised devices necessitate a robust and distributed revocation scheme to ensure the integrity and trustworthiness of the system~\cite{satybaldyTaxonomy2024}. Existing SSI frameworks address aspects of decentralized IdM but often lack mechanisms to allow flexible onboarding of individual issuers or to dynamically manage trust propagation across complex IoT networks. These conditions underscore the need for a decentralized framework that empowers users to issue credentials while maintaining a secure chain of trust through endorsement-based mechanisms, thereby reducing dependency on manufacturers and centralized entities~\cite{kaariainenBuilding2024}.

\textbf{Contributions.} In this paper, we propose a blockchain-based SSI framework for IoT, which addresses critical challenges in decentralized IdM, including flexible credential issuance, trust propagation, and secure credential revocation. Our layered framework enables any individual with a verifiable trust linkage to trusted issuers (e.g., manufacturers) to become an issuer, ensuring scalability and decentralization. To establish trust, we employ a hierarchical trust model that calculates endorsement-based trust scores, allowing the system to dynamically onboard issuers and securely manage trust relationships. We incorporate blockchain as a Verifiable Data Registry (VDR) to ensure transparency and immutability of identity-related operations, while smart contracts automate core SSI processes, including credential issuance, verification, and revocation. The framework improves privacy by decoupling pseudonymous and sensitive identity records, and enhances user autonomy by enabling device owners to manage their credentials without reliance on centralized entities. We implement our proposed framework as a proof-of-concept on a permissioned blockchain network, demonstrating its feasibility and efficiency in meeting the requirements of IoT ecosystems.


\textbf{Related work.}
The application of SSI in IoT has been extensively explored in the literature, focusing on various dimensions of IdM and trust~\cite{muktaCredTrust2022,dayaratneSSI4IoT2024,fathallaLightweight2024,fathallaPTSSIM2023}. Mukta et al.~\cite{muktaCredTrust2022} introduced \textit{CredTrust}, a blockchain-based SSI framework enabling individual users to act as credential issuers. The model employs a hierarchical trust propagation method, allowing official issuers to onboard individual issuers by endorsing their credentials. While CredTrust demonstrates feasibility for managing issuer trust through trust propagation, its dependency on explicit endorsements from official issuers limits the scalability of onboarding in more dynamic IoT scenarios. Dayaratne et al.~\cite{dayaratneSSI4IoT2024} addressed challenges in applying SSI to IoT by proposing optimizations for verifiable credentials (VCs), such as reducing their size and life-cycle management in constrained IoT environments. However, their framework primarily supports credential issuance by manufacturers or predefined trusted entities, restricting flexibility for users to issue credentials for their devices. Fathalla and Azab~\cite{fathallaLightweight2024} proposed \textit{SSIoT}, a lightweight SSI framework designed for zero-trust IoT networks. By integrating secret sharing schemes and blockchain, the model secures identity credentials and ensures minimal computational overhead for resource-constrained IoT devices. Nevertheless, SSIoT lacks mechanisms to decentralize credential issuance to individuals beyond cloud service agents, thus maintaining a strong reliance on centralized control. Unlike prior work, our proposed framework emphasizes a flexible and decentralized trust model, enabling any individual with a verifiable trust linkage to official issuers (e.g., manufacturers) to onboard as credential issuers. This approach extends the scope of SSI for IoT by addressing limitations in hierarchical trust propagation and enabling decentralized credential issuance, improving scalability and usability in dynamic IoT ecosystems.

\section{Proposed System Model} 
\label{sec:system-model}
Our proposed model involves multiple entities, each playing distinct roles to ensure secure and decentralized IdM in IoT, which are described in the following subsections.

\subsection{Entities and Their Corresponding Roles}
\label{sub:actors}
Our proposed SSI model, denoted $\mathbb{SSI} = (M, U, D)$, defines three primary entities, where $M=\{m_1,\ldots,m_k\}$ is the set of IoT manufacturers, $U=\{u_1,\ldots,u_l\}$ is the set of users, and $D=\{d_1,\ldots,d_p\}$ is the set of IoT devices. We classify the devices into two categories: \textit{strong devices} and \textit{weak devices}, denoted $\mathcal{D}^{p}_{type} \subseteq \{\mathit{strong}, \mathit{weak}\}$. The roles and responsibilities of these entities are described as follows:

\textbf{IoT Manufacturers.}
The manufacturers $M$ are responsible for issuing initial credentials to IoT devices during production. These credentials are anchored to a decentralized trust registry, i.e., the blockchain layer (cf. Sec.~\ref{sub:layer-view}), to ensure their authenticity and immutability. Manufacturers correspond as the official issuer to which individual user anchor their trustworthiness to issue new credentials.

\textbf{Users.}
The users $U$ are the primary controllers of their devices within the $\mathbb{SSI}$ framework, which are responsible to manage device credentials, and transfer ownership. Users have the authority to bind weak devices to strong devices within their control and revoke credentials for compromised devices. This enables dynamic management of IoT networks while minimizing dependency on centralized intermediaries.

\textbf{IoT Devices.}
The IoT devices $D$ are categorized as strong or weak:
\begin{itemize}
    \item \textit{Strong Devices}: Devices in this category, denoted $\mathcal{D}^{p}_{type} = \mathit{strong}$, have higher computational capabilities for performing SSI processes. These devices delegate operations for weak devices, ensuring efficient communication and security. Examples include smartphones, IoT hubs, or gateway devices.
    \item \textit{Weak Devices}: As opposed to their stronger counterpart, these devices are resource-constrained and rely on strong devices for authentication and credential management. These devices typically perform specific tasks, such as sensing or actuation, and benefit from the delegated trust established by their associated strong devices.
\end{itemize}
The devices in the network is uniquely mapped to its owner $u_l$ within the $\mathbb{SSI}$ framework. In the context of SSI, devices can act as a holder and verifier.

\begin{figure}
    \centering
    \includegraphics[width=\linewidth]{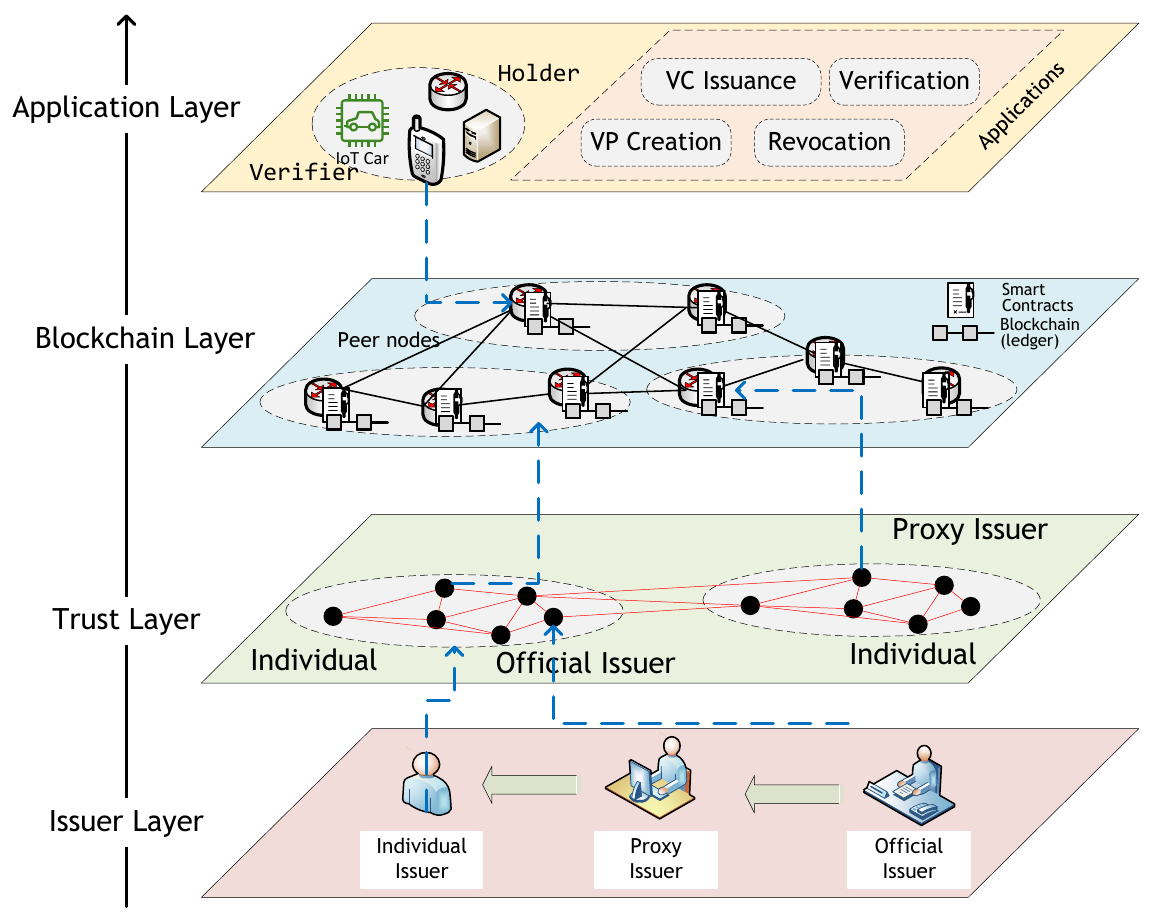}
    \caption{The proposed SSI ecosystem model, $\mathbb{SSI}$, with its distinct layers. These layers allow for each individual to be trusted issuers for their IoT devices.}
    \label{fig:layer-diagram}
\end{figure}

\subsection{Layered View of the System}
\label{sub:layer-view}
Fig.~\ref{fig:layer-diagram} presents a layered view of $\mathbb{SSI}$. We categorize the layers into four interconnected layers, denoted $\mathbb{SSI} = \left<\mathbb{L}_{app}, \mathbb{L}_{bc}, \mathbb{L}_{trust}, \mathbb{L}_{isr}\right>$, namely application, blockchain, trust, and issuer layers.

\textbf{Issuer layer $\mathbb{L}_{isr}$.}
This layer is responsible for onboarding issuers into the SSI system. It ensures that each individual issuer has a verifiable connection to the root of trust, represented by official issuers such as IoT manufacturers. To achieve this, $\mathbb{L}_{isr}$ introduces proxy issuers, which act as intermediaries connecting individual issuers to the official issuers. Proxy issuers extend the hierarchical chain of trust, allowing flexible and scalable onboarding while maintaining a secure link to the root of trust. Each onboarded issuer is associated with a decentralized identifier (DID) and is added to the network's VDR, ensuring their authenticity and traceability.

\textbf{Trust layer $\mathbb{L}_{trust}$.}
This layer provides the abstraction of the web of trust, inspired by the PGP trust model~\cite{garfinkel1995pgp}. The trust layer manages the relationships between all onboarded issuers and their connections to the official issuers. It ensures that individual issuers are only onboarded if their trustworthiness can be verified through endorsements or relationships with the root issuers. By maintaining a dynamic web of trust~\cite{garfinkel1995pgp}, this layer mitigates the risk of malicious or unknown entities acting as issuers. The web of trust also facilitates the propagation of trust throughout the system, enabling secure credential issuance and revocation~\cite{muktaCredTrust2022}.

\textbf{Blockchain layer $\mathbb{L}_{bc}$.}
The blockchain layer serves as the operational backbone of the $\mathbb{SSI}$ model. A blockchain network acts as the VDR, ensuring that all identity-related data, such as credential issuance and revocation records, is transparent, tamper-proof, and immutable. Smart contracts within this layer automate critical SSI processes, including credential issuance, trust calculations, and revocation updates~\cite{dharmaputraDeTRM2022}. These smart contracts enforce the business logic of the SSI system, ensuring compliance with the trust and issuer layers while enabling interoperability with existing systems.

\textbf{Application layer $\mathbb{L}_{app}$.}
This layer provides an interface for end-users and applications to interact with the SSI system. It supports a wide range of SSI functionalities, such as authentication and credential verification. Applications in this layer communicate directly with the blockchain layer to query or update credential information.

\subsection{Onboarding Individual Issuers}
\label{sub:onboarding}
In our model, onboarding new individual issuers into $\mathbb{SSI}$ relies on a hierarchical chain-of-trust mechanism~\cite{garfinkel1995pgp}. This mechanism ensures that trust propagates from manufacturers, who act as the root of trust, to all connected individual issuers. By leveraging endorsement vectors and decentralized trust anchors, the system allows anyone with a direct or indirect link to a trusted issuer to securely become an onboarded issuer.

\textbf{Endorsement-based trust calculation.}
Each issuer in the system is associated with an endorsement score, which quantifies their trustworthiness based on connections to trusted issuers (e.g., manufacturers). Let the set of all issuers $\mathbb{I} = (M, U)$, where $\mathbb{I} = \{i_1, i_2, \ldots, i_n\}$, and let each issuer $i_j$ have a set of endorsements $E_{j} = \{e_{j,1}, e_{j,2}, \ldots, e_{j,m}\}$. Each endorsement $e_{j,k}$ is a scalar value representing the level of trust provided by another issuer or the manufacturer. The trust score ${T}_{i_j}$ for an issuer $i_j$ is calculated as:
\begin{equation}
\label{eq:issuer-trust}
    {T}_{i_j} = \frac{1}{|E_j|} \sum_{k=1}^{|E_j|} w_k e_{j,k},
\end{equation}
where $|E_j|$ is the cardinality of $E_j$ and $w_k$ is the weight assigned to each endorsement based on the trustworthiness of the endorsing entity. The weights are derived from the endorsers' trust scores, ensuring that endorsements from more trusted entities have a greater influence.

\textbf{Propagation of trust.}
The hierarchical model ensures that trust propagates through the network via direct or indirect connections. If an individual issuer $i_j$ lacks a direct endorsement from a manufacturer, their trust score can still be established through intermediate issuers in the web of trust. For example, let $i_j$ be indirectly connected to a manufacturer via a chain of issuers $i_{x_1}, i_{x_2}, \ldots, i_{x_k}$, where each intermediate issuer provides an endorsement to the next. The trust score for $i_j$ is updated iteratively along the chain:
\begin{equation}
\label{eq:trust-propagation}
    {T}_{i_j} = \prod_{x=1}^{k} {T}_{i_{x}},
\end{equation}
where ${T}_{i_{x}}$ is the trust score of each intermediate issuer in the chain.

\textbf{Onboarding process.}
An individual issuer $i_j$ is successfully onboarded if their calculated trust score ${T}_{i_j}$ meets a predefined threshold $\tau$, i.e., ${T}_{i_j} \geq \tau$. This threshold ensures that only sufficiently trusted issuers are allowed to issue credentials within the system. Once onboarded, the issuer is assigned a DID and is registered in the VDR maintained on the blockchain layer. 

\textbf{Proxy issuers.}
To facilitate onboarding in cases where direct connections to manufacturers are impractical, the system supports proxy issuers. Proxy issuers are trusted intermediaries that endorse new issuers on behalf of the manufacturers. Each proxy issuer operates under strict trust requirements and is accountable for the endorsements they provide. This approach decentralizes the onboarding process while maintaining the integrity of the chain-of-trust.

\begin{figure*}
    \centering
    \begin{tabularx}{\linewidth}{XXX}
        \includegraphics[width=0.32\textwidth]{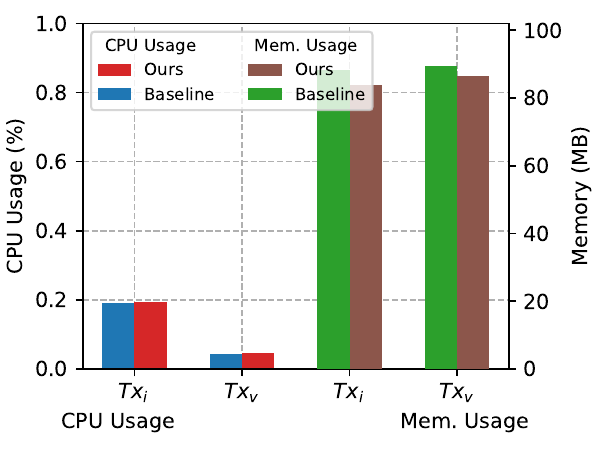}
        \caption{Comparison of CPU and memory consumption for issue and verify transactions between our solution and the baseline.}
        \label{fig:cpu-mem-usage}
        &
        \includegraphics[width=0.32\textwidth]{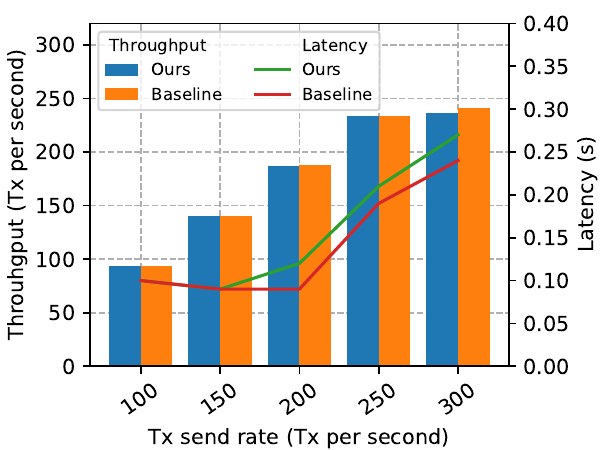}
        \caption{Comparison of throughput and latencies for issue transactions with varying number of transaction send rates.}
        \label{fig:tx-tps}
        &
        \includegraphics[width=0.32\textwidth]{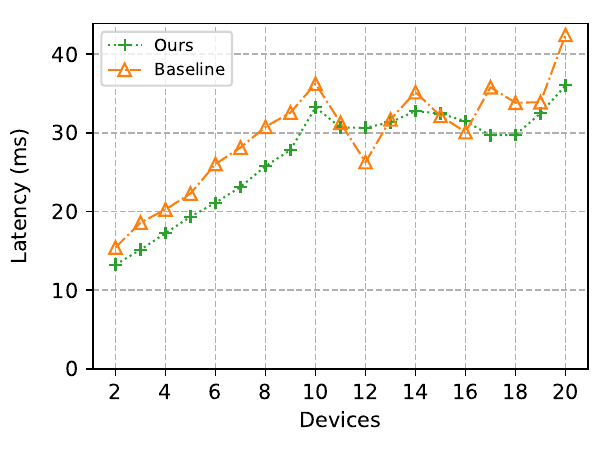}
        \caption{Authentication latency under varying numbers of parallel requests, which shows relatively minor fluctuations with stable values.}
        \label{fig:latency}
    \end{tabularx}
\end{figure*}

\section{Blockchain-based SSI Framework}
\label{sec:blockchain-ssi-framework}
To implement the proposed $\mathbb{SSI}$, we leverage a blockchain-based framework that ensures transparency, immutability, and decentralized control of identity-related operations. In $\mathbb{L}_{bc}$, the blockchain $\mathbb{B}$ acts as the VDR for storing credential issuance, verification, and revocation records~\cite{paikDecentralizedIdentityManagement2024}. The key blockchain transactions in our model include $\mathsf{onboard}$, $\mathsf{issue}$, $\mathsf{verify}$, and $\mathsf{revoke}$, each described below.

$\mathsf{onboard}$ \textbf{transaction.}
To join the issuer network, an individual issuer $i_j$ must provide their trust linkage to a trusted issuer (e.g., manufacturer). The blockchain evaluates the linkage to determine if there exists a direct or indirect connection to a manufacturer. If such a connection exists, the individual is deemed trusted and onboarded into the system. The $\mathsf{onboard}$ transaction is defined as:
\begin{equation}
\label{eq:tx_onboarding}
    Tx_o = [\, Tx^{o}_{ID} \parallel L(i_j) \parallel \mathit{timestamp} \parallel Sig_{i_j} \,],
\end{equation}
where $Tx^{o}_{ID}$ is the unique onboarding transaction ID, $L(i_j)$ is the trust linkage of $i_j$, and $Sig_{i_j}$ is the signature of the individual issuer. Upon successful onboarding, $i_j$ is a DID and registered in $\mathbb{B}$.

$\mathsf{issue}$ \textbf{transaction.}
To issue a credential, an issuer must invoke the $\mathsf{issue}$ transaction. The blockchain validates whether the issuer has a valid trust connection to a trusted or official issuer. If the trust linkage is verified, the credential is issued, resulting in a Verifiable Credential (VC) assigned to the holder. The $\mathsf{issue}$ transaction is defined as:
\begin{equation}
\label{eq:tx_issue}
    Tx_i = [\, Tx^{i}_{ID} \parallel C_j \parallel \mathit{timestamp} \parallel Sig_{i_j} \,],
\end{equation}
where $C_j=\{c_{j,1}, c_{j,2}, \ldots, c_{j,n}\}$ represents $n$ claims $c_{j,n}=\left<key,val\right>$ in key-value pair of the associated properties, which also includes the holder's DID. The blockchain ensures that the issuer is authorized to issue the credential based on their trust score.

$\mathsf{verify}$ \textbf{transaction.}
To validate a Verifiable Credential (VC), a verifier invokes the $\mathsf{verify}$ transaction. The blockchain checks the validity of the VC by verifying its cryptographic signature and ensuring it has not been revoked. The $\mathsf{verify}$ transaction is defined as:
\begin{equation}
\label{eq:tx_verification}
    Tx_v = [\, Tx^{v}_{ID} \parallel VC_i \parallel \mathit{timestamp} \parallel Sig_{v} \,],
\end{equation}
where $Tx^{v}_{ID}$ is the transaction ID, and $Sig_v$ is the verifier's signature. If the verification succeeds, the VC is confirmed as valid.

$\mathsf{revoke}$ \textbf{transaction.}
To revoke credentials for stolen or compromised devices, an issuer invokes the $\mathsf{revoke}$ transaction. The blockchain updates the status of the associated VC, marking it as revoked. The $\mathsf{revoke}$ transaction is defined as:
\begin{equation}
\label{eq:tx_revocation}
    Tx_r = [\, Tx^{r}_{ID} \parallel VC_i \parallel R_i \parallel \mathit{timestamp} \parallel Sig_{i_j} \,],
\end{equation}
where $R_i$ specifies the rationale for revocation (e.g., "compromised" or "stolen"). The blockchain ensures that the issuer invoking the revocation has the authority to do so.

The blockchain framework enables seamless interaction between these transactions to ensure decentralized IdM required for IoT applications~\cite{fedrecheskiSelfSovereign2020,samirDTSSIM2022}.
For example, an individual issuer must first complete an $\mathsf{onboard}$ transaction to join the network. Once onboarded, the issuer can invoke $\mathsf{issue}$ transactions to create credentials for their devices. Verifiers can validate these credentials via $\mathsf{verify}$ transactions, while compromised credentials can be invalidated using $\mathsf{revoke}$ transactions.

\section{Proof of Concept Evaluation} 
\label{sec:poc-evaluation}
We developed a proof-of-concept implementation of the proposed SSI framework on Hyperledger Fabric v2.5, incorporating chaincodes to implement $\mathsf{onboard}$, $\mathsf{issue}$, $\mathsf{verify}$, $\mathsf{revoke}$ transactions, and endorsement-based trust management. The evaluation compares our system against a baseline SSI framework, which lacks the endorsement-based mechanisms and limits credential issuance to trusted manufacturers.
We first investigated the CPU and memory consumption of our framework under normal operations. The experiment compared the resource usage with the baseline SSI system to evaluate the additional overhead introduced by our solution. As shown in Fig.~\ref{fig:cpu-mem-usage}, while there is a slight increase in CPU in our framework, the overhead is negligible. This result demonstrates that the proposed endorsement-based mechanisms can be implemented without significantly impacting the computational efficiency of the underlying blockchain system.
Next, we evaluated the throughput of our framework during the execution of credential issuance transactions ($Tx_i$). The experiment measured the number of transactions successfully processed per second and compared the results with the baseline system. As depicted in Fig.~\ref{fig:tx-tps}, our framework achieves relatively similar throughput to the baseline, indicating that the introduction of endorsement-based trust calculations does not degrade the system's ability to process transactions efficiently.
Lastly, we assessed the latency of authentication operations in a simulated IoT interaction scenario. In this experiment, multiple parallel authentication requests were issued, and the latency was recorded as the number of simultaneous requests increased. Fig.~\ref{fig:latency} illustrates that while there are minor fluctuations, our latency is relatively similar to the baseline. These results confirm that the proposed framework maintains responsiveness, even under higher loads.
Overall, our evaluation demonstrates that the proposed SSI framework introduces only minimal overheads compared to the baseline. By maintaining comparable CPU and memory usage, achieving relatively similar throughput, and incurring negligible latency, our framework proves to be well-suited for dynamic and resource-constrained IoT environments.

\section{Conclusion}
In this paper, we proposed a blockchain-based SSI framework for IoT, addressing challenges in decentralized credential issuance, trust propagation, and revocation. The framework employs a layered architecture where trust is dynamically established through endorsement-based mechanisms and maintained using a hierarchical chain-of-trust model. Experimental results indicate that the system is feasible and incurs minimal overheads, while providing scalability.
By leveraging blockchain as the VDR and automating critical processes with smart contracts, our framework ensures decentralized IdM suitable for dynamic IoT ecosystems.

\section*{Acknowledgment}
\noindent The authors acknowledge the support from the Department of Electrical and Information Engineering UGM. This work was also supported by Hibah Penelitian Fundamental (PFR), Ministry of Higher Education, Science, and Technology Indonesia, grant number 048/E5/PG.02.00.PL/2024 - 2679/UN1/DITLIT/PT.01.03/2024.


\bibliographystyle{IEEEtran}
\bibliography{./biblio}

\end{document}